\documentclass[intlimits,twoside,a4paper]{article}

\usepackage[cp1251]{inputenc}

\usepackage{subcaption}
\usepackage[eqsecnum]{cmpj3}

\usepackage{array,multirow,makecell}
\usepackage{color}

\issue{2025}{28}{1}{13701}
\doinumber{10.5488/CMP.28.13701}
\title[Thermoelectric properties of $\textrm{CeIr}_\textrm{4} \textrm{P}_\textrm{12}$]%
{Theoretical study of thermoelectric properties of $\textrm{CeIr}_\textrm{4} \textrm{P}_\textrm{12}$ filled skutterudite for
energy conversion }

\author[M. Bouchenaki, L. I. Karaouz\`{e}ne, B. N. Brahmi, M. Kaid Slimane]{ M. Bouchenaki\orcid{0009-0009-9038-2092}\thanks{Corresponding author: \email{manel.bouchenaki@univ-tlemcen.dz}, \email{bouchenaki.manel@gmail.com}}, L. I. Karaouz\`{e}ne\orcid{0000-0002-2971-2939},  
        B. N. Brahmi\orcid{0000-0003-4576-0241}, M. Kaid Slimane\orcid{0009-0000-1011-0025}}
\address{Theoretical Physics Laboratory, Faculty of Science, University Abou Bekr Belkaid, Tlemcen, Algeria}


\date{Received January 21, 2025, in final form February 18, 2025}

\begin{document}

\maketitle

\begin{abstract}
The structural, elastic, thermodynamic and thermoelectric characteristics of the $\textrm{CeIr}_\textrm{4} \textrm{P}_\textrm{12}$ skutterudite have been predicted for the first time by applying density functional theory and the semi-classical Boltzmann simulations. Firstly, the structural-magnetic stability was verified through ground-state energy calculations obtained from structural optimizations. The predicted single-crystal elastic constants (${C}_{ij}$) show that the title compound is mechanically stable. At the same time, it turns out to be dynamically stable where all the calculated phonon frequencies have positive values. The cohesive energy was calculated to verify the energy stability of the material considered. We also determined the variations of some macroscopic physical parameters as functions of temperature, namely the thermal expansion coefficient, the lattice thermal conductivity. Furthermore, we investigated the temperature dependencies of some thermoelectric coefficients such as electronic thermal conductivity, and figure of merit. Such encouraging results indicate that the compound is a potential candidate for thermoelectric devices.
\printkeywords
\end{abstract}

\section{Introduction}

With an increase in energy and environmental concerns, the search for renewable energy sources and the development of high-efficiency energy conversion technologies has become a major issue in our societies. A wide range of alternative energy technologies such as solar and wind energy have been developed, especially thermoelectric technology, which is why thermoelectric materials were widely studied in the past decades \cite{andrea,kashy}.
Thermoelectric devices are currently used in various applications, such as automotive thermoelectric generators~\cite{shen2019}, in space (NASA multi-mission radioisotope thermoelectric generator)~\cite{woerner2016,holgate2015}, solar thermoelectric generators \cite{karthick2019}, and other technologies~\cite{jaziri2020,champier2017}. They offer such benefits as easy configuration, longer lifespan, absence of vibrations, noise, and respect for the environment~\cite{champier2017,he2018}.
However, thermoelectric devices remain reserved for commercial use due to their low efficiency compared to other types of electricity generators \cite{shang2021,zoui2020}. It is therefore essential to discover effective methods of improving the performance of the materials used for the construction of thermoelectric devices \cite{gayner2016}.
Thermoelectric material efficiency is defined by the figure of merit \emph{ZT}, which relies on the Seebeck coefficient \emph{S}, electrical conductivity $\sigma$, thermal conductivity $\kappa$ and absolute temperature \emph{T}. This dimensionless number is given by the following relation \cite{zair2023,zair2022,goldsmid2010}:
\begin{equation}
    ZT=\frac{S^2\sigma}{\kappa}T.
\end{equation}

Among the various thermoelectric materials we can find skutterudites. The origin of the name skutterudite derives from its place of finding, namely the cobalt mines in Skutterud (Norway), where the naturally occurring mineral $\textrm{CoAs}_\textrm{3}$ was firstly discovered and described in 1845 \cite{wei2014,breithaupt1827}.
They are typically opaque with a metallic luster and tin-white to silver-gray in color. Today, as an accessory mineral, they are found in many localities worldwide \cite{benhalima2021optimisation}.
The general formula of binary skutterudite is $\textrm{MX}_\textrm{3}$ (or more opportune: $\textrm{M}_\textrm{4}\textrm{X}_\textrm{12}$), where M is a transition metal (Co, Fe, Ir and Rh) and X is a pnictogen atom (As, Sb and P). The crystal structure of the $\textrm{MX}_\textrm{3}$ binary skutterudites was established in 1928 by Oftedal \cite{oftedal1928xxxiii}.
According to him, the atomic arrangement in skutterudite classified the mineral $\textrm{CoAs}_\textrm{3}$ in space group $Im\Bar{3}$ (204). 
Despite the high Seebeck coefficient and high electrical conductivity of binary skutterudites, their thermal conductivity is extremely high.
The structure of binary skutterudite includes interstitial sites that can accommodate atoms such as alkalins or rare earths to form ternary filled skutterudites $\textrm{RM}_\textrm{4}\textrm{X}_\textrm{12}$, thereby further improving the figure of merit by reducing thermal conductivity \cite{bashir2020,liu2019new,qin2020,masarrat2021,al2021,jiang2020,nolas1998}. Indeed, filled skutterudites are affordable thermoelectric materials that exhibit low thermal conductivity, resulting in an improved thermoelectric performance.

Numerous studies on materials based on rare earth filled skutterudites structures have revealed that such materials possess a variety of intriguing thermoelectric properties. Sanada et al. \cite{sanada2005} found an additional strong evidence that the filled skutterudite structure has unusual electronic states, where multi $f$ electron ions can also show unprecedented strongly correlated electron behavior. Chaki~et~al.~\cite{chaki2021first} calculated and analyzed the thermoelectric properties of the filled skutterudite $\textrm{SmRu}_\textrm{4}\textrm{Sb}_\textrm{12}$. They found a high Seebeck coeficient and a figure of merit ($ZT$) equal to 0.38 at 300~K. Furthermore, Shankar~et~al.~\cite{shankar2017} investigated the electronic structure and the thermal transport properties of the filled skutterudite $\textrm{EuRu}_\textrm{4}\textrm{As}_\textrm{12}$. The authors report that the material has  semi-metallic ferromagnetic nature and the compound has a figure of merit ($ZT$) value of 0.55.
Abdelakader et al. \cite{abdelakader2024} explored the thermoelectric properties of filled skutterudites $\textrm{ThFe}_\textrm{4}\textrm{P}_\textrm{12}$ and $\textrm{CeFe}_\textrm{4}\textrm{P}_\textrm{12}$ using DFT calculations and reported that the highest recorded figure of merit values for both materials is 0.14 for $\textrm{ThFe}_\textrm{4}\textrm{P}_\textrm{12}$ and 0.10 for $\textrm{CeFe}_\textrm{4}\textrm{P}_\textrm{12}$ at 900 K.

To our knowledge, there have not been any previous theoretical calculations of the transport properties in Ir-P based rare earth full-filled skutterudites. In this paper, we are interested in the full-filled skutterudite $\textrm{CeIr}_\textrm{4}\textrm{P}_\textrm{12}$ compound by studying the structural, elastic, thermodynamic and thermoelecric properties using the full potential linearized augmented plane wave (FP-LAPW) methods. Due to a lack of any previous studies on this sample we believe that the present theoretical investigation will fill the gap that exists in Ir-P based rare earth full-filled skutterudites and will be a guide to future experimental endeavors.
This paper is arranged as follows. We present the calculation procedure
in section \ref{section2}. Next, our results are detailed in section \ref{section3}, in which the different structural, electronic, thermodynamic and thermoelectric properties will be determined and discussed. The conclusion is given in section \ref{section4}, where
we summarize our main findings.

\section{Computational details}
\label{section2}
The structural properties were determined by performing the first-principle calculations which are based on density functional theory (DFT) with the full potential linearized augmented plane wave (FP-LAPW) method implemented in the WIEN2k code \cite{blaha2001wien2k,sayah2021,zeffane2021,adnane2020}. The exchange-correlation potential was calculated with the parameterization of the generalized gradient approximation (GGA) by Wu and Cohen (GGA-WC)~\cite{wu2006} and local density approximation (LDA) \cite{becke1993}. To treat the effect of strongly correlated $f$-orbital on the electronic and hence on the thermoelectric properties, we applied a Hubbard $U$ correction for Ce only, since it has an $f$-orbital which corresponds to a value of $U= 7.0$~eV and $J= 0.7$~eV \cite{calderon2015}.
In the framework of the FP-LAPW method, to solve the Kohn--Sham equation, the unit cell was divided into non-overlapping spheres, known as Muffin-tin spheres. The muffin-tin radii {$R_{MT}$} for each element were chosen to be equal to 2.5, 2.26 and 1.85 Bohr for Ce, Ir and P, respectively. We used a parameter $R_{MT}k_{\rm max}=8$, which determines the matrix size, where $k_{\rm max}$ is the plane wave cut-off and $R_{MT}$ is the smallest of all atomic radii. The maximum plane waves used is $G_{\rm max}=14$. The core and valence-states-energy-gap (cut-off energy) \color{black} is chosen to be $-6$~Ry. Checking the dynamic stability was done through phonon band dispersion and phonon DOS analysis of both compounds using the PHONOPY package~\cite{togo2015}. Effect temperature and pressure on thermodynamic properties such as heat capacity, Debye temperature, etc. are estimated using the quasi-harmonic approximation implemented in GIBBS2 code~\cite{otero2011,otero2011gibbs2}. On the other hand, the thermal transport properties have been calculated using Boltzmann semi-classical theory~\cite{allen1996b} as implemented in the BoltzTraP2 code \cite{madsen2018}.

\section{Results and discussion}
\label{section3}
\subsection{ Structural and magnetic properties}
\begin{figure}[ht]
   \centering
   \includegraphics[scale=0.9]{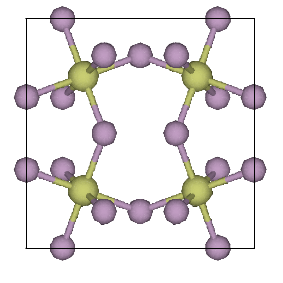}
   \includegraphics[scale=0.9]{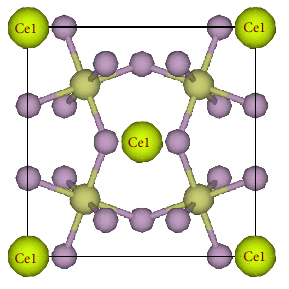}
   \caption{(Colour online) Unit cell of binary (a) and Ce-filled skutterudite (b).}
   \label{figure1}
\end{figure}

The $\textrm{IrP}_\textrm{3}$ is a binary skutterudite crystallized in the centered cubic structure, the atom Ir occupies crystallographic positions $8c$ of the unit cell (1/4, 1/4, 1/4), the P atoms occupy the crystallographic positions $24g$ $(0, u, v)$. The centered cubic cell contains 32 atoms. The filler atom Ce is located at the $2a$ (0, 0, 0) Wyckoff positions, see figure \ref{figure1}.

Since we are interested in the properties of $\textrm{CeIr}_\textrm{4}\textrm{P}_\textrm{12}$ compound, a careful identification of their equilibrium structure is needed. We calculated the total energy of the system for different unit cell volumes, and the equilibrium lattice constant was obtained by fitting the calculated data to the Murnaghan equation of state \cite{murnaghan1944}.  
The internal free parameters ($u$ and $v$) have been optimized by minimizing the total energy while keeping the volume fixed.
The results of the structural optimization, namely the equilibrium lattice constant $a_0$, internal parameters $u$ and $v$, the bulk modulus
$B$ and its first pressure derivative $B'$ are listed in table~\ref{table1}.

\begin{table}[h]
    \centering
    \medskip
    \caption{Calculated structural parameters for binary $\textrm{IrP}_\textrm{3}$ and filled skutterudite $\textrm{CeIr}_\textrm{4}\textrm{P}_\textrm{12}$.}
    \begin{tabular}{ccccccccc}
    \hline
    \small Materials &  & &\small $a_0 (\mathring{A}) $ &\small $u$ &\small $v$ &\small $B$ (GPa) &\small $B'$  &\small $E_0$(Ry) \\ 
    \hline
    \small $\textrm{IrP}_\textrm{3}$ &\footnotesize Experimental \cite{nolas1999skutterudites} &\small \multirow{3}{2em}{NM} &\small 8.015 & $-$ & $-$& $-$ & $-$ & $-$ \\
      &\footnotesize Other work (GGA-PBE) \cite{benhalima2021optimisation} & & \small 8.109 &\small 0.354 &\small 0.139 &\small 177.290 &\small 4.551 &\small $-151076.237$ \\
      &\footnotesize Our calculation (GGA-WC) & &\small 8.054 &\small 0.357 & \small 0.137 &\small 184.129 &\small 4.733 &\small $-151061.214$ \\
      \hline
      \small $\textrm{CeIr}_\textrm{4}\textrm{P}_\textrm{12}$ & \footnotesize \multirow{2}{4em}{GGA-WC}  &\small FM &\small 8.238 &\small 0.340 &\small 0.158 &\small 163.864 &\small 4.749 &\small $-168790.136$ \\
      &  &\small NM  &\small 8.250 &  &  &\small 164.248 &\small 4.644 &\small $-168790.104$ \\
      &\footnotesize LDA &\small FM &\small 8.197 &  &  &\small  176.824 &\small 5.481 &\small $-168687.265$ \\
      \hline
    \end{tabular}
    
    \label{table1}
\end{table}

\begin{figure}[ht]
           \centering
               \begin{subfigure}[b]{0.45\textwidth}
               \includegraphics[width=\textwidth]{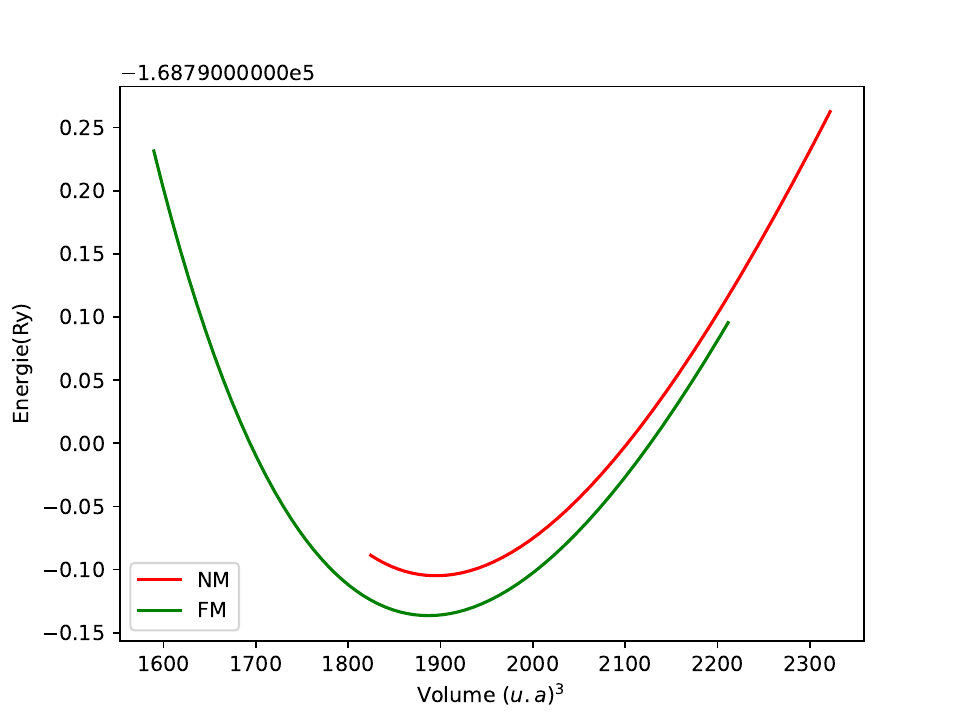}
               \caption{}
               \label{figure2a}
               \end{subfigure}
               \begin{subfigure}[b]{0.45\textwidth}
               \includegraphics[width=\textwidth]{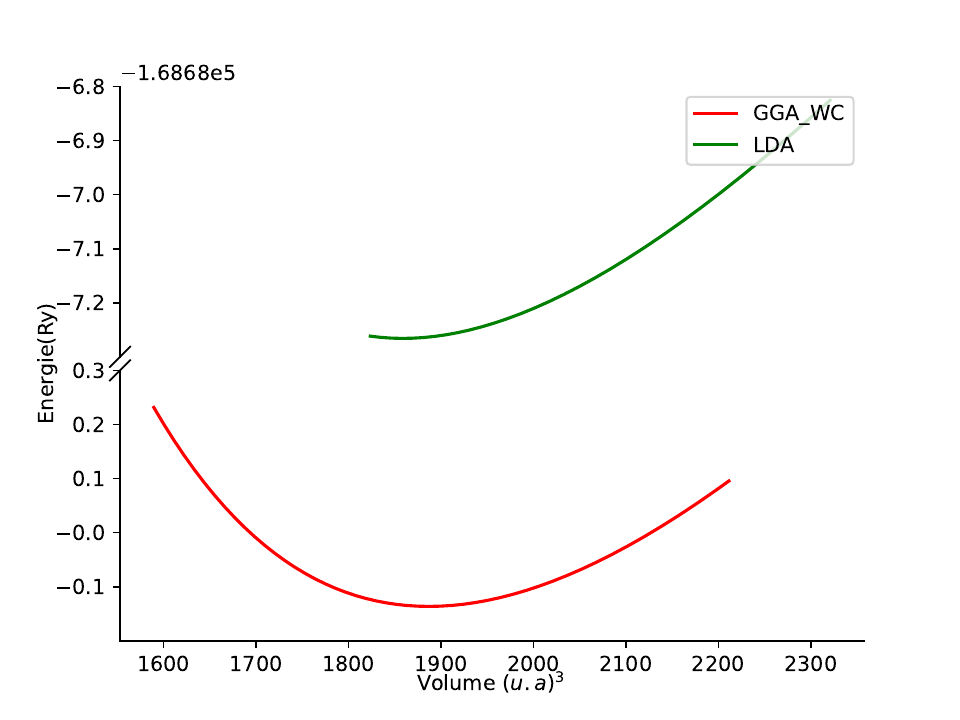}
               \caption{}
               \label{figure2b}
               \end{subfigure}
           \caption{(Colour online) The variation of the energies versus the volume of $\textrm{CeIr}_\textrm{4}\textrm{P}_\textrm{12}$  (a) in FM and PM States, (b) by LDA and GGA-WC.}
           \label{figure2}
\end{figure}

Figure \ref{figure2a} shows the total energy-volume curves for $\textrm{CeIr}_\textrm{4}\textrm{P}_\textrm{12}$ compound in  FM (ferromagnetic) and PM (paramagnetic) states. Our results exhibit that the compound with ferromagnetic arrangement have the lowest energies at the equilibrium state. Thus, they are energetically more stable compared to paramagnetic.

We can note that there is no experimental and theoretical calculation of the structural parameters to compare with our results. Thus, in order to consolidate our predictive results, we used a comparative study between the two approximations used in our work to predict an experimental value of the lattice parameters~$a_0$. In table~\ref{table1}, we can compare the results of the structural properties obtained by LDA (increase) and GGA-WC (decrease) approximations with ferromagnetic arrangement in order to predict the range of experimental lattice parameter $a_0$ (see figure \ref{figure2b}). It can be seen that the lattice constant of $\textrm{CeIr}_\textrm{4}\textrm{P}_\textrm{12}$ is larger than that of $\textrm{IrP}_\textrm{3}$. This increase in the lattice parameter indicates the expansion of the crystal due to the filling of rare earths in the voids of $\textrm{IrP}_\textrm{3}$.

The calculated local and total magnetic moments using GGA, GGA+U and LDA of $\textrm{CeIr}_\textrm{4}\textrm{P}_\textrm{12}$ are listed in table \ref{table2}.

\begin{table}[h]
    \centering
    \medskip
    \caption{Calculated total and atomic magnetic moment in terms of $\mu_{\text{B}}$ of $\textrm{CeIr}_\textrm{4}\textrm{P}_\textrm{12}$.}
    \begin{tabular}{ccccccc}
    \hline
     \multirow{2}{*}{} & \multirow{2}{*}{Approximations} & Interstial magnetic  & \multicolumn{3}{c}{Atomic magnetic moment}  &  Total magnetic  \\ 
    \cline{4-6}
     &  & moment & Ce & Ir & P & moment \\
     \hline
     \multirow{3}{*}{$\textrm{CeIr}_\textrm{4}\textrm{P}_\textrm{12}$} & GGA-WC & 0.55470 & 1.10098 & 0.05726 & 0.02808 & 2.22174 \\
     & GGA-WC+U & 0.56614 & 1.72740 & 0.10801 & 0.01912 & 2.95501 \\
     & LDA & 0.45294 & 0.93713 & 0.03418 & 0.02128 & 1.78221 \\
     \hline

    \end{tabular}
    
    \label{table2}

\end{table}

\subsection{Stability}
As our work is a predictive study, we are obliged to see the stability of our compound $\textrm{CeIr}_\textrm{4}\textrm{P}_\textrm{12}$.

\subsubsection{Thermodynamic stability}
The thermodynamic stability of this compound can be predicted by calculating the formation energy, which can be calculated by the following equation: 
\begin{equation}
    E_f=E_{\rm CeIr_{4}P_{12}}-(E_{\rm Ce}+4E_{\rm Ir}+12E_{\rm P}),
\end{equation}
where $E_{\rm CeIr_{4}P_{12}}$ is the total fundamental energy of the compound $\textrm{CeIr}_\textrm{4}\textrm{P}_\textrm{12}$ per formula unit, $E_{\rm Ce}$, $E_{\rm Ir}$, and~$E_{\rm P}$ are the total fundamental energies of the bulk crystals of Ce, Ir, and P, respectively. The calculated formation energy has a negative value ($-6.82201891$~Ry) which indicates the thermodynamic stability of the structure.

\subsubsection{Mechanical stability}
To confirm the mechanical stability of our compounds, we should calculate only three independent elastic parameters $C_{11}$, $C_{12}$ and $C_{44}$ due to cubic symmetry of the compound from the second-order derivatives of the fitted polynomials of the total energy.  Fulfilment of the stability criteria \cite{born1996}, $C_{11}+2C_{12} > 0$, $C_{44} >0$ and $C_{11}-C_{12} > 0$ suggests that the compound is stable under shape deformation. We computed several parameters such as the bulk and shear modulus, Young's modulus ($Y$), Poisson's ratio ($\nu$) and the anisotropy constant ($A$). The obtained results using Voigt Reuss-Hill approximations (VRH) \cite{voigt1908,reuss1929,hill1952} are summarized in table \ref{tabel3}. We note that to date, no experimental or previous theoretical results are available to be compared with our obtained results. The mathematical relationship between the elastic constants and each of these elastic parameters is given as follows:
\begin{equation}
    B=\frac{1}{3}(C_{11}+2C_{12}),
\end{equation}
\begin{equation}
    G_{v}=\frac{1}{5}(C_{11}+3C_{44}-C_{12}),
\end{equation}
\begin{equation}
    Y=\frac{9BG_{v}}{3B+G_{v}},
\end{equation}
\begin{equation}
    A=\frac{2C_{44}}{C_{11}-C_{12}}.
\end{equation}

The bulk modulus is a measure of the ability of a substance to withstand the changes in volume when under compression on all sides, the results obtained by using Voigt Reuss-Hill approximations are equal.
Visibly, the bulk moduli calculated from elastic constants are in good agreement with those provided by the EOS (table \ref{table1}).
The shear modulus $G$ describes the response of the material to shear stress. The strength of a material can be quantified by its Pugh ratio \cite{pugh195}.
This ratio refers to the nature of the material: ductile if ($B/G$) greater than 1.75 and brittle in the reverse case. Pugh's ratio values obtained indicate that they possess ductile nature with high malleability. Young's modulus $Y$ describes the strain response of the material to uniaxial stress in the direction of this stress. The Poisson's ratio ($\nu$) indicates the nature of bonding and stiffness of the material, the Poisson rate is related to the nature of
the bonds between the atoms where for typical covalent bonding its value is  around 0.1 whereas for typical ionic crystals it is  up to 0.25~\cite{haines2001}. Thus, according to our results, Poisson's coefficients are greater than 0.25, and the dominant bond  for the compound is ionic. The anisotropic factor $A$ is found to be different from 1 which shows that the material is anisotropic in nature.

\begin{table}[h]
	\centering
	\medskip
	\caption{Calculated mechanical parameters for $\textrm{CeIr}_\textrm{4}\textrm{P}_\textrm{12}$.}
	\begin{tabular}{ccc}
		\hline
		\multirow{3}{*}{Elastic constants} & $C_{11}$ (GPa) & 212.1808 \\
		& $C_{12}$ (GPa) & 149.0571 \\
		& $C_{44}$ (GPa) & 35.9092 \\
		\hline
		Bulk modulus & $B$ (GPa) & 170.098 \\
		\hline
		\multirow{3}{*}{Shear modulus} & $G_{V}$ (GPa) & 34.193 \\
		& $G_{R}$ (GPa) & 34.055 \\
		& $G_{H}$ (GPa) & 34.124 \\
		\hline
		\multirow{3}{*}{Young modulus} & $Y_{V}$ (GPa) & 96.137 \\
		& $Y_{R}$ (GPa) & 95.773 \\
		& $Y_{H}$ (GPa) & 95.955 \\
		\hline
		\multirow{3}{*}{Poisson's ratio} & $\nu_{V}$ & 0.405 \\
		& $\nu_{R}$  & 0.406 \\
		& $\nu_{H}$  & 0.405 \\
		\hline
		Anisotropy ratio & $A$ (J/m$^{3}$) & 1.1377 \\
		\hline
		Pugh's ratio & $B/G$ & 4.974 \\
		\hline
		
	\end{tabular}
	\label{tabel3}
\end{table}

\subsubsection{Dynamic stability}
\begin{figure}
    \centering
    \includegraphics[width=0.7\linewidth]{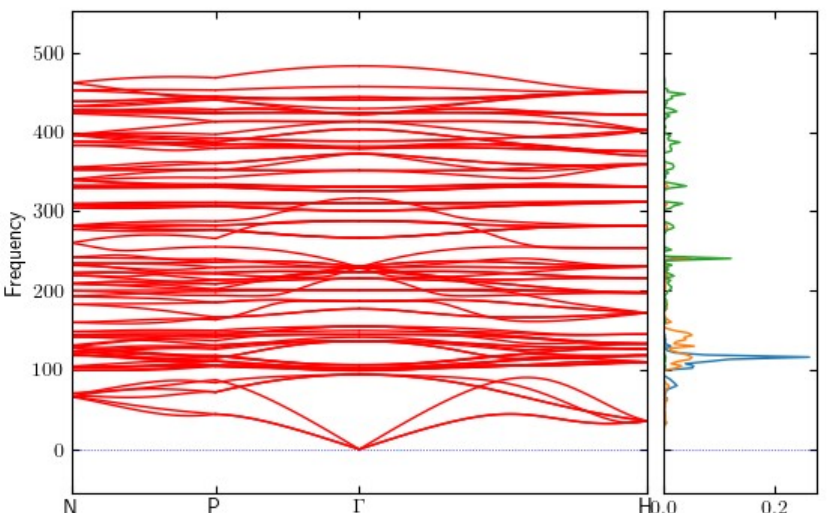}
    \caption{(Colour online) Phonon dispersion and phonon density of states of the $\textrm{CeIr}_\textrm{4}\textrm{P}_\textrm{12}$ skutterudite. }
    \label{figure3}
\end{figure}

Figure \ref{figure3} shows the phonon dispersion spectra, total and projected density of phonon states of the compound $\textrm{CeIr}_\textrm{4}\textrm{P}_\textrm{12}$ at a pressure of 0 GPa. Analysis of band phonon spectra shows that the compound has a shape with no space separating the acoustic and optical modes, and we can see that all phonon branches (whether acoustic or optical) have positive frequencies at zero pressure. Obviously, this indicates the dynamic stability of these compounds.

\subsection{Electronic properties}
\begin{figure}[ht]
           \centering
               \begin{subfigure}[b]{0.43\textwidth}
               \includegraphics[width=\textwidth]{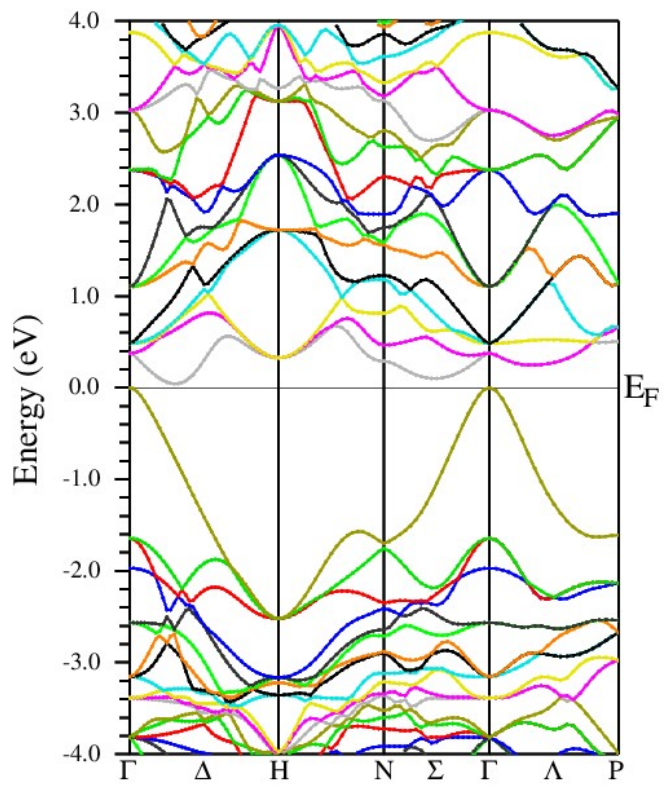}
               \caption{}
               \label{figure4a}
               \end{subfigure}
               \begin{subfigure}[b]{0.53\textwidth}
               \includegraphics[width=\textwidth]{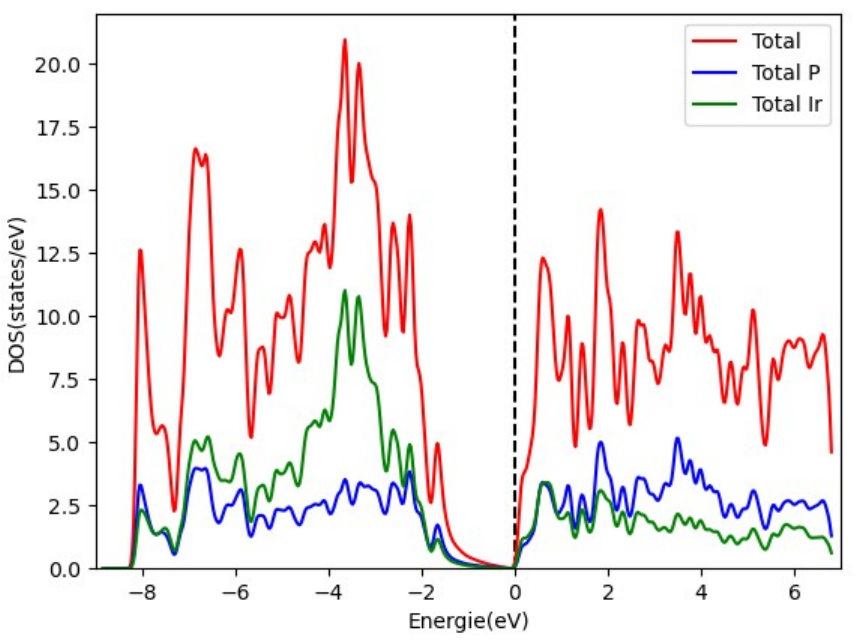}
               \caption{}
               \label{figure4b}
               \end{subfigure}
           \caption{(Colour online) (a) Band structure and (b) density of states for $\textrm{Ir}\textrm{P}_\textrm{3}$.}
           \label{figure4}
\end{figure}

We have computed the band structures of binary and Ce-filled $\textrm{Ir}_\textrm{4}\textrm{P}_\textrm{12}$ to investigate the electrical characteristics of the current material. The electronic band structures of the binary skutterudite is shown in figure \ref{figure4}. The maximum of the valence band is located at point $\Gamma$ and the minimum of the conduction band is located at point H, we conclude that $\textrm{Ir}\textrm{P}_\textrm{3}$ has an indirect gap of 0.376 eV along the $\Gamma$--H direction, which agrees
well with \cite{benhalima2021optimisation}.

\begin{figure}[ht]
           \centering
               \begin{subfigure}[b]{0.4\textwidth}
               \includegraphics[width=\textwidth]{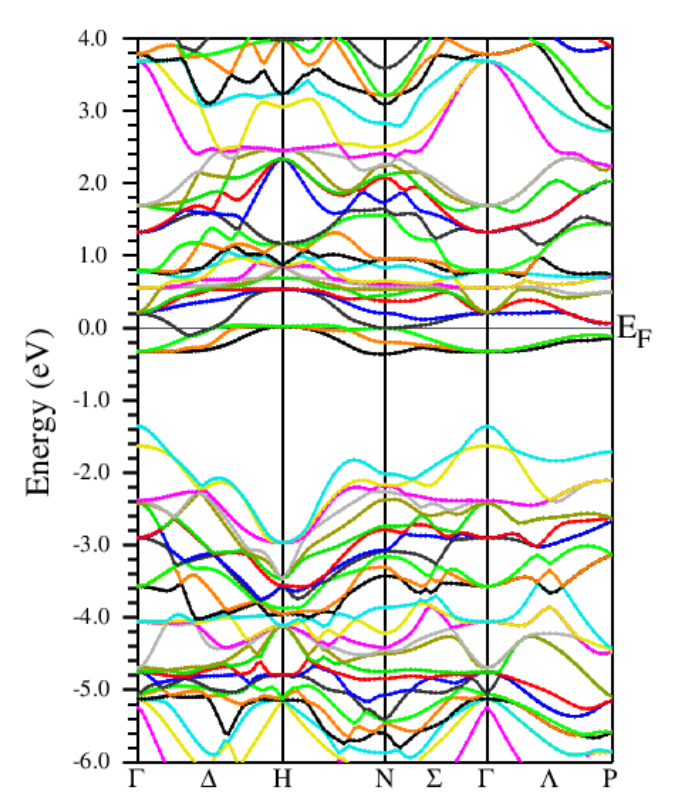}
               \caption{}
               \label{figure5a}
               \end{subfigure}
               \begin{subfigure}[b]{0.4\textwidth}
               \includegraphics[width=\textwidth]{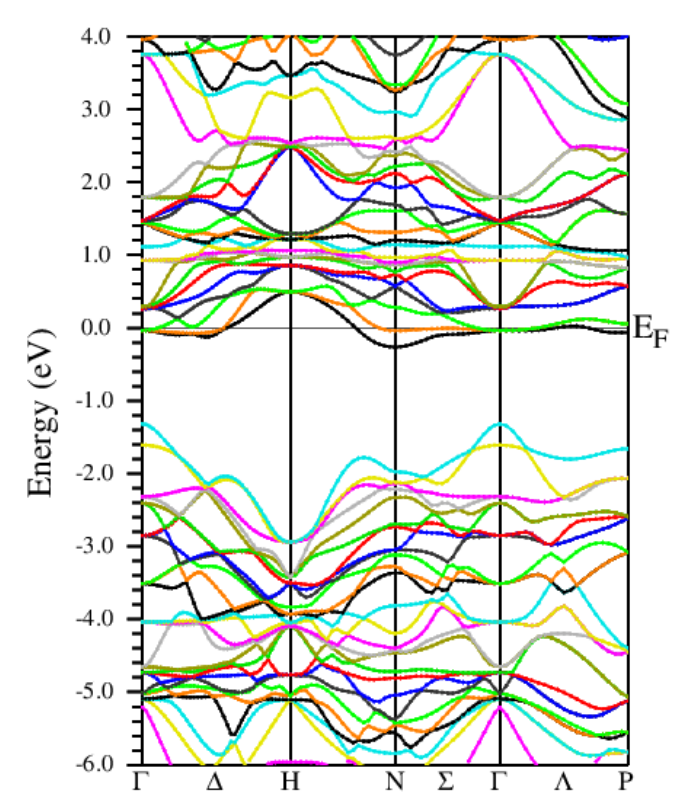}
               \caption{}
               \label{figure5b}
               \end{subfigure}
           \caption{(Colour online) Electronic band structure of $\textrm{CeIr}_\textrm{4}\textrm{P}_\textrm{12}$, (a) spin up, (b) spin down.}
           \label{figure5}
\end{figure}

The energy band structures of $\textrm{CeIr}_\textrm{4}\textrm{P}_\textrm{12}$ along the high symmetry directions of body centered cubic Brillouin zone (BZ) are shown in figure \ref{figure5} for spin up and spin down channels. It is evident that the Fermi energy level is not in between the valance and conduction band region. Additionally, it is discovered that three bands that emerge from the conduction band's bottom pass the Fermi level and enter the valance band in both channels. This implies that the substance has a metallic character.

\begin{figure}[ht]
	\centering
	\begin{subfigure}[b]{0.45\textwidth}
		\includegraphics[width=\textwidth]{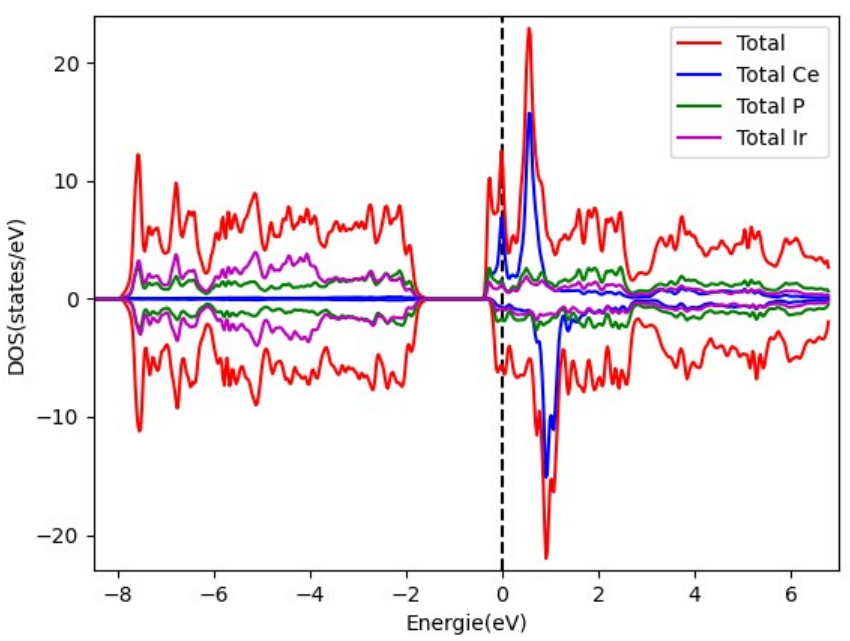}
		\caption{}
		\label{figure6a}
	\end{subfigure}
	\begin{subfigure}[b]{0.45\textwidth}
		\includegraphics[width=\textwidth]{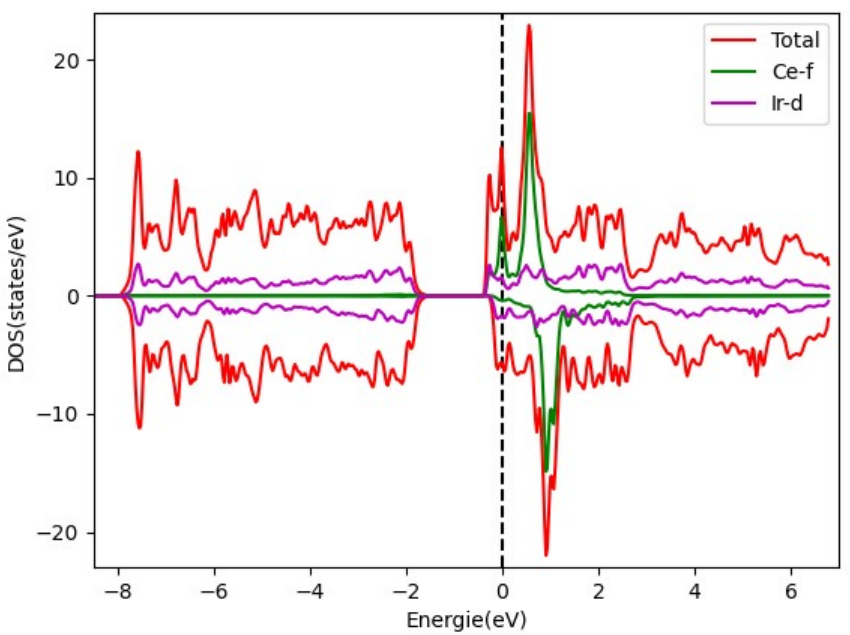}
		\caption{}
		\label{figure6b}
	\end{subfigure}
	\caption{(Colour online) (a) Total DOS and (b) partial DOS for $\textrm{CeIr}_\textrm{4}\textrm{P}_\textrm{12}$.}
	\label{figure6}
\end{figure}

Additionally, for both spin up and spin down electrons, we have computed the total and partial density of states (DOS). These findings are shown in figure~\ref{figure6}.

From the analysis of PDOS plot, we observed that for both spin channels, the Ce $f$-states dominate in the valence band. The contribution of other orbitals is negligible compared to that of $4f$-Ce electronic state.
The study of DOS plots also indicates the metallic nature of the given ternary skutterudite compound. Since no data are available in literature related to the electronic properties of $\textrm{CeIr}_\textrm{4}\textrm{P}_\textrm{12}$, no comparison is given here.

\begin{figure}
    \centering
    \includegraphics[width=0.6\linewidth]{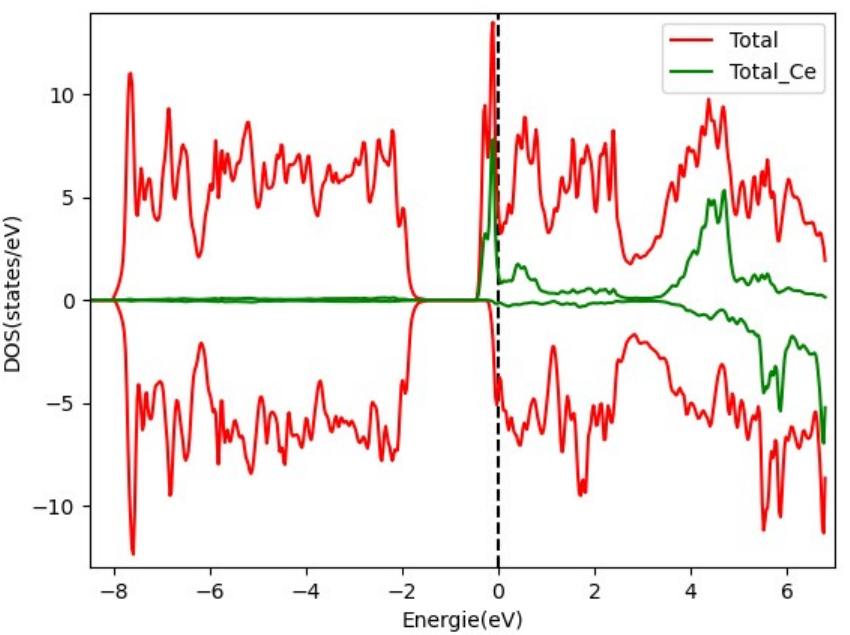}
    \caption{(Colour online) Density of states with Hubbard parameter for $\textrm{CeIr}_\textrm{4}\textrm{P}_\textrm{12}$. }
    \label{figure7}
\end{figure}

The effect of the Hubbard parameter on the density of states plot of $\textrm{CeIr}_\textrm{4}\textrm{P}_\textrm{12}$ skutterudite is clear
from the plots shown in figure \ref{figure7},  where we can see it in the magnetic properties in table \ref{table2}.  

\subsection{Thermodynamic properties}
The calculations obtained by using the WIEN2k program are performed at 0 K temperature, meaning that we neglected the vibrational side of the atoms in the hamiltonian equation of the crystal system with a view to
simplify the Schr\"odinger equation and this was via applying 
the Born--Oppenheimer approximation \cite{born1985}. For this reason, the effect of temperature as well as the thermal properties are investigated using the quasi-harmonic model \cite{debye1926einige} implemented in GIBBS2 code \cite{otero2011,otero2011gibbs2}.

\begin{figure}[!ht]
           \centering
               \begin{subfigure}[b]{0.45\textwidth}
               \includegraphics[width=\textwidth]{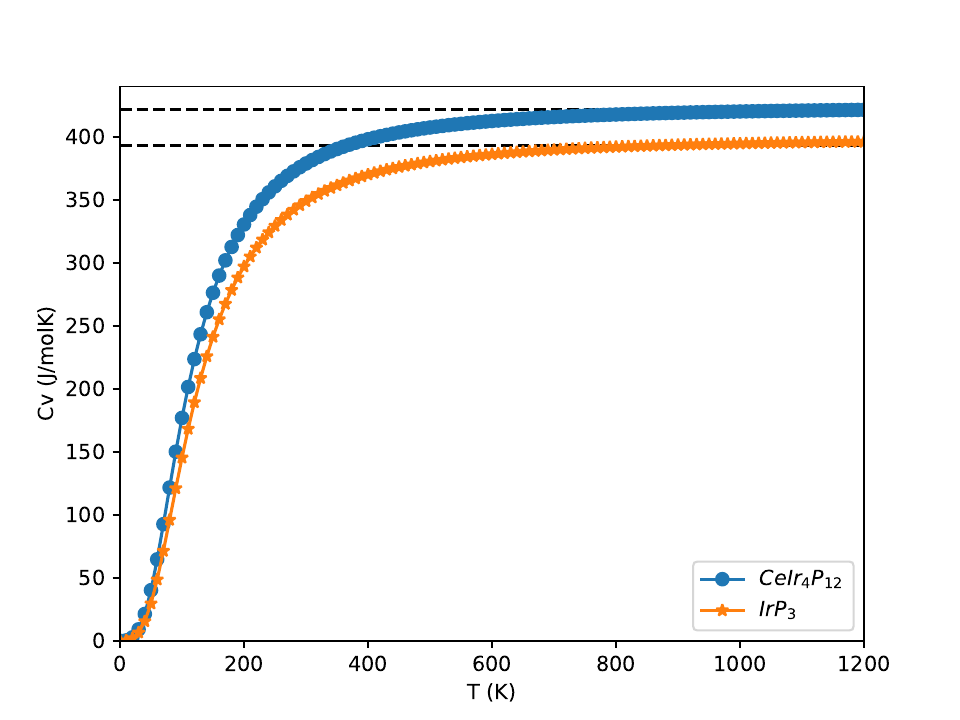}
               \caption{}
               \label{figure8a}
               \end{subfigure}
               \begin{subfigure}[b]{0.45\textwidth}
               \includegraphics[width=\textwidth]{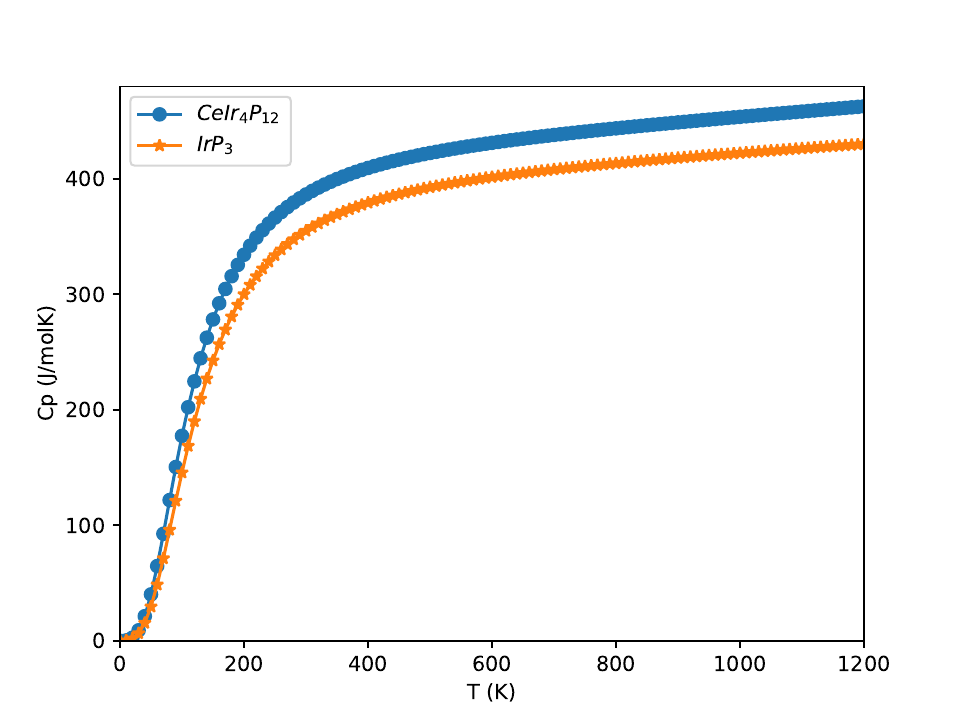}
               \caption{}
               \label{figure8b}
               \end{subfigure}
               \begin{subfigure}[b]{0.45\textwidth}
               \includegraphics[width=\textwidth]{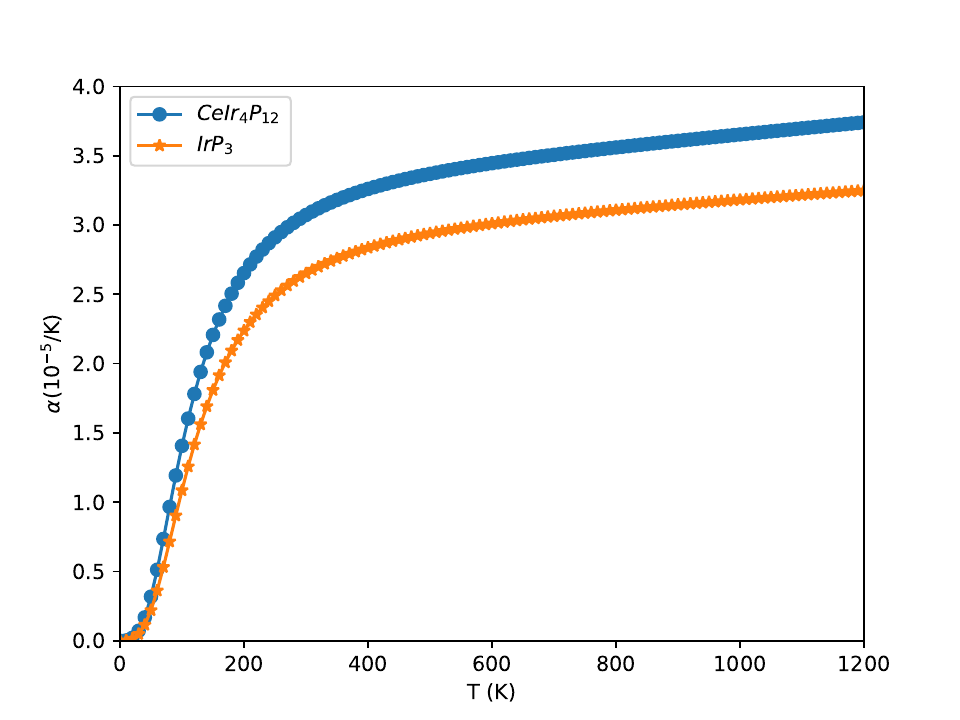}
               \caption{}
               \label{figure8c}
               \end{subfigure}
               \begin{subfigure}[b]{0.45\textwidth}
               \includegraphics[width=\textwidth]{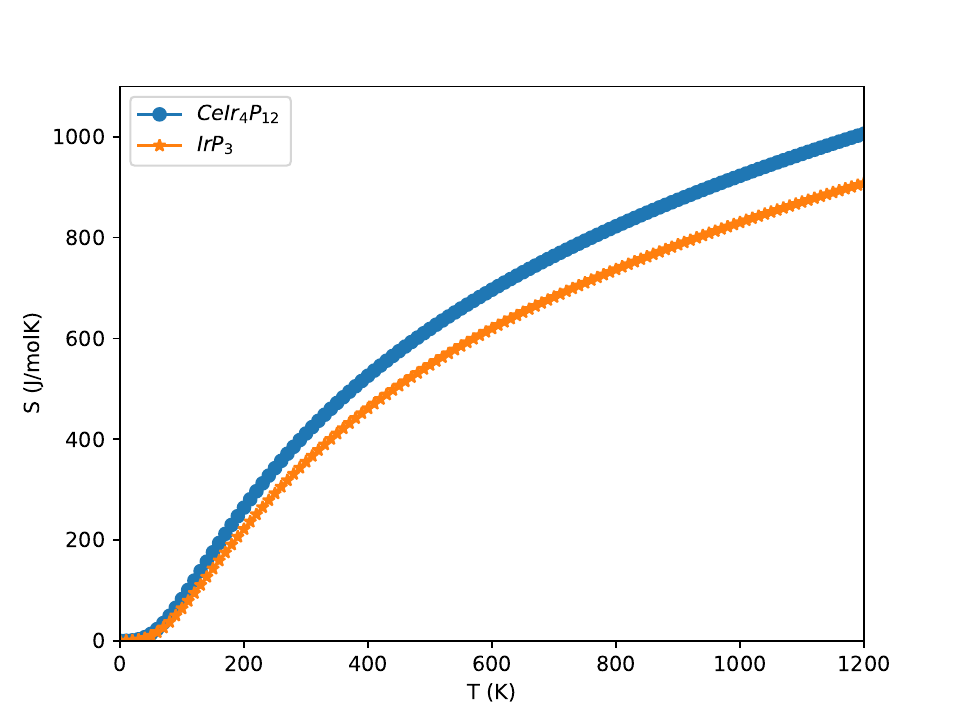}
               \caption{}
               \label{figure8d}
               \end{subfigure}
               \begin{subfigure}[b]{0.45\textwidth}
               \includegraphics[width=\textwidth]{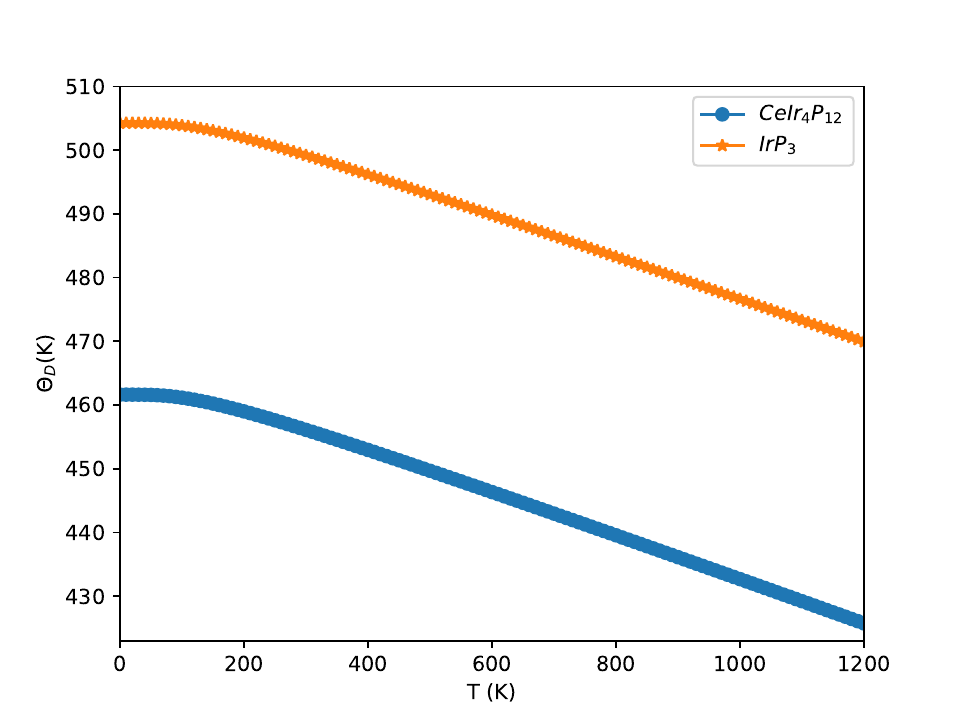}
               \caption{}
               \label{figure8e}
               \end{subfigure}
               \begin{subfigure}[b]{0.45\textwidth}
               \includegraphics[width=\textwidth]{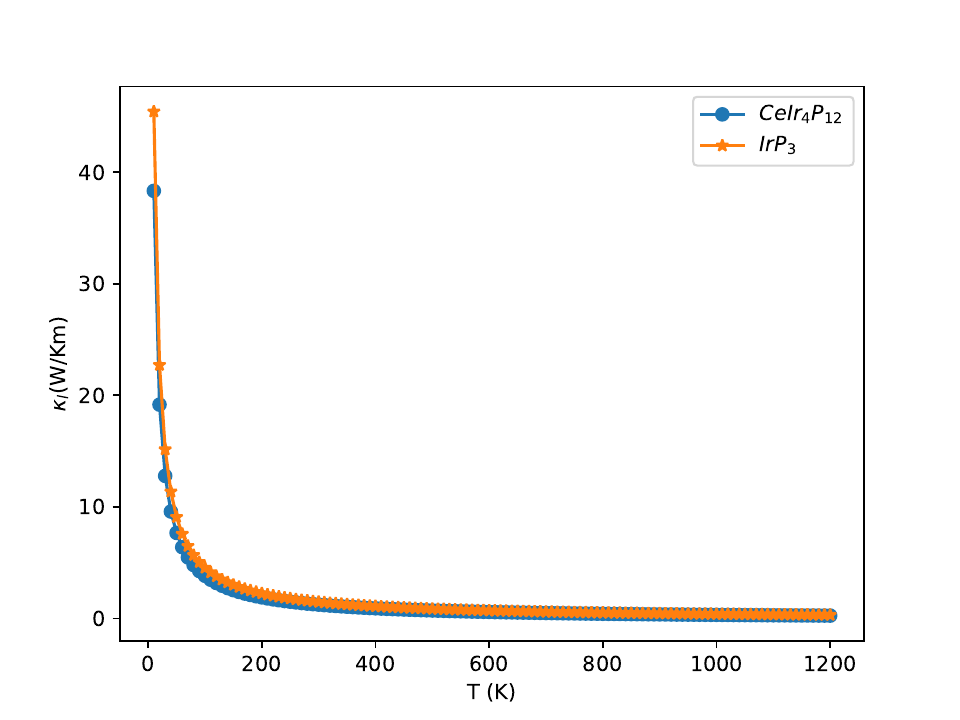}
               \caption{}
               \label{figure8f}
               \end{subfigure}
           \caption{(Colour online) Variation of heat capacities [(a) $C_V$ and (b) $C_P$], (c) thermal expansion ($\alpha$), (d) entropy ($S$), (e) Debye temperature ($\Theta_D$) and (f) crystal conductivity ($\kappa_l$) of $\textrm{Ir}\textrm{P}_\textrm{3}$ and $\textrm{CeIr}_\textrm{4}\textrm{P}_\textrm{12}$ as a function of temperature.}
           \label{figure8}
\end{figure}

The temperature effect on heat capacities at constant volume and pressure, thermal expansion, entropy, Debye temperature and thermal conductivity in the temperature range of 0 to 1200 K are shown in figure~\ref{figure8}.

The heat capacity at constant volume (see figure~\ref{figure8a}) checked the Dulong-Petit law at high temperatures \cite{dulong1819} ,where its value reached 393 and 422~J~mol$^{-1}$K$^{-1}$ for $\textrm{Ir}\textrm{P}_\textrm{3}$ and $\textrm{CeIr}_\textrm{4}\textrm{P}_\textrm{12}$, respectively when the temperature is greater than 550 K, while $C_V$ is proportional to $T^3$ at low temperature. Meanwhile, we can see that the heat capacity at constant pressure (see figure~\ref{figure8b}) has the behavior similar to the capacity~$C_V$.

The variation of the coefficient of thermal expansion ($\alpha$) as a function of temperature is shown in figure~\ref{figure8c}. According to this figure, we can see that the thermal expansion increases very rapidly for temperatures smaller than 250~K, and it increases slowly with temperatures greater than 250~K and becomes almost linear.  Figure~\ref{figure8d} shows that the entropy increases with an increase of the temperature because the increase of the temperature leads to a growth of the modes of vibration and consequently the number of configurations possible. Debye temperature is the most important thermal property. It is associated with many mechanical and elastic properties and it represents the corresponding temperature which caused the highest possible number of the modes vibration. Through figure~\ref{figure8e}, we note that the temperature of Debye varies inversely with the change in temperature. In order to know the distribution of the contributions of phonons in the thermal conductivity, we studied in figure~\ref{figure8f} the changes in crystal conductivity $\kappa_l$ in terms of temperature. The $\kappa_l$ can be calculated using the Slack model formula given by~\cite{slack1973}:
\begin{equation}
    \kappa_l=\frac{AM\Theta_{\rm D}^3\delta}{\gamma^2Tn^{\frac{2}{3}}},
\end{equation}
where $A$ is a physical constant equal to:
\begin{equation}
    A=\frac{2.43\cdot 10^{-18}\gamma^2}{\gamma^2-0.514\gamma+0.228}.
\end{equation}
$\Theta_{\rm D}$, $\gamma$, $\delta^3$, $n$ and $M$ are the Debye temperature, Gr\"uneisen parameter, the volume per atom, the number of atoms in the primitive unit cell and the average mass of all the atoms in the crystal, respectively.

Figure \ref{figure8f} shows that $\kappa_l$ exponentially decreases with an  increasing temperature and reaches its lowest level when the temperature is in the range greater than 500 K.

\subsection{Thermoelectric properties}
To study how the waste heat from our samples can be converted into beneficial electrical energy we measured the thermoelectric transport properties. For this purpose, we used BoltzTraP2 code interfaced
to WIEN2k in order to  calculate  the requisite transport quantities, such as electrical conductivity ($\sigma$), Seebeck coefficient~($S$), thermal conductivity ($\kappa$), and dimensionless thermoelectric figure of merit ($ZT$) as a
function of temperature of binnary skutterudite $\textrm{Ir}\textrm{P}_\textrm{3}$ and filled skutterudite $\textrm{CeIr}_\textrm{4}\textrm{P}_\textrm{12}$ in the temperature range 100--1200 K, for chemical potentials values $\mu_0= 0.725656$~eV, $\mu_0= 0.791948$~eV and $\mu_0= 0.791951$~eV of $\textrm{Ir}\textrm{P}_\textrm{3}$, $\textrm{CeIr}_\textrm{4}\textrm{P}_\textrm{12}$ spin up and $\textrm{CeIr}_\textrm{4}\textrm{P}_\textrm{12}$ spin down, respectively (these are the closest values to the Fermi level). We note that for Ir-P based rare earth filled skutterudites, and we are not aware of any theoretical works where the temperature dependence of thermoelectric properties is reported.

\begin{figure}[ht]
    \centering
    \includegraphics[width=0.55\linewidth]{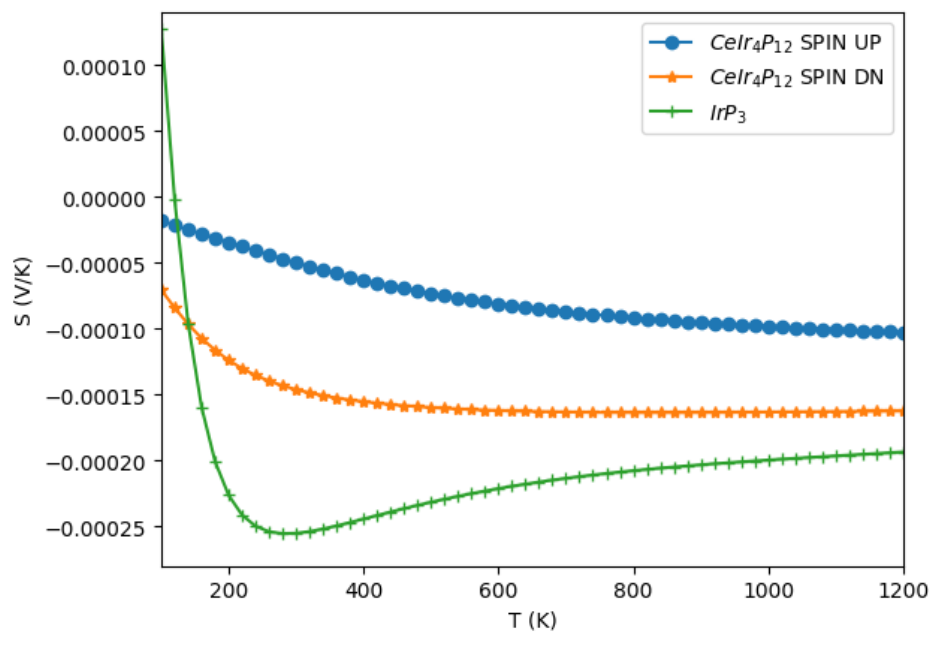}
    \caption{(Colour online) Seebeck coefficient ($S$) in V/K as a function of temperature for $\textrm{Ir}\textrm{P}_\textrm{3}$ and $\textrm{CeIr}_\textrm{4}\textrm{P}_\textrm{12}$ spin up and spin down.}
    \label{figure9}
\end{figure}

The evolution of the Seebeck coefficient with the temperature, calculated for $\textrm{Ir}\textrm{P}_\textrm{3}$ and $\textrm{CeIr}_\textrm{4}\textrm{P}_\textrm{12}$, is shown in figure~\ref{figure9}. For $\textrm{Ir}\textrm{P}_\textrm{3}$, the value of $S$ drops to a minimum at room temperature and then begins to increase with an increasing temperature. For $\textrm{CeIr}_\textrm{4}\textrm{P}_\textrm{12}$ the value of $S$ monotonously decreases with an increasing temperature while still being negative, demonstrating its $n$-type character in spin up as well as in spin down. The differentiation between $p$-type and $n$-type characteristics in various spin states is essential for the development of spintronic thermoelectric devices.

\begin{figure}[ht]
    \centering
    \includegraphics[width=0.55\linewidth]{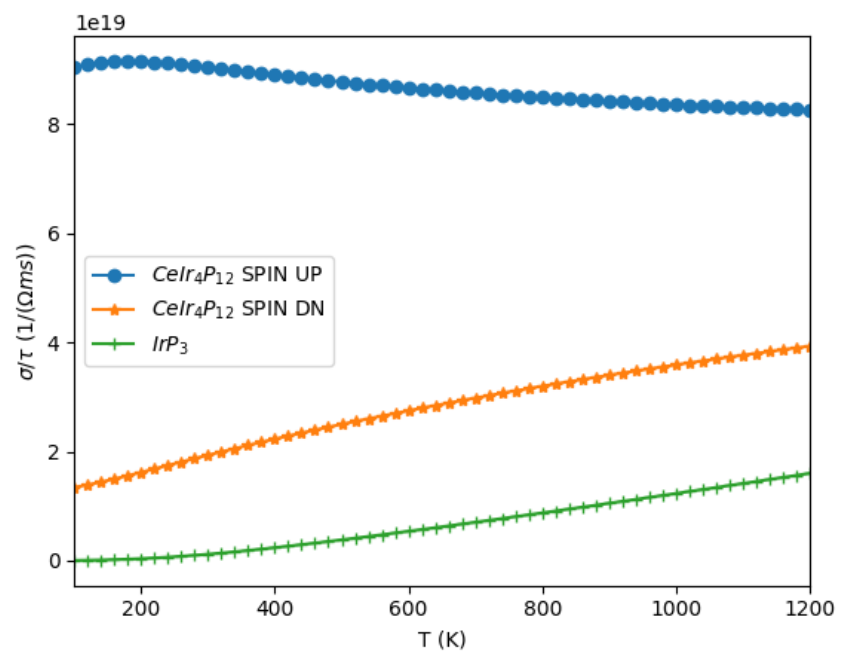}
    \caption{(Colour online) Electric conductivity ($\sigma/\tau$) in $\Omega $~ms$^{-1}$ as a function of temperature for $\textrm{Ir}\textrm{P}_\textrm{3}$ and $\textrm{CeIr}_\textrm{4}\textrm{P}_\textrm{12}$ spin up and spin down.}
    \label{figure10}
\end{figure}

Electric conductivity ($\sigma/\tau$) refers to the capacity of a substance to carry electric current. Figure~\ref{figure10} illustrates the behavior of the electrical conductivity of  $\textrm{Ir}\textrm{P}_\textrm{3}$ and $\textrm{CeIr}_\textrm{4}\textrm{P}_\textrm{12}$. Upon increasing temperature, the electrical conductivity has a slightly decreasing aspect for the spin up case of $\textrm{CeIr}_\textrm{4}\textrm{P}_\textrm{12}$. In the spin down state and for $\textrm{Ir}\textrm{P}_\textrm{3}$, the value of ($\sigma/\tau$) grows drastically, reaching about $4\times 10^{19}$ and $1.6 \times 10^{19}$~$\Omega$~ms$^{-1}$ at 1200 K, respectively.

\begin{figure}[ht]
    \centering
    \includegraphics[width=0.55\linewidth]{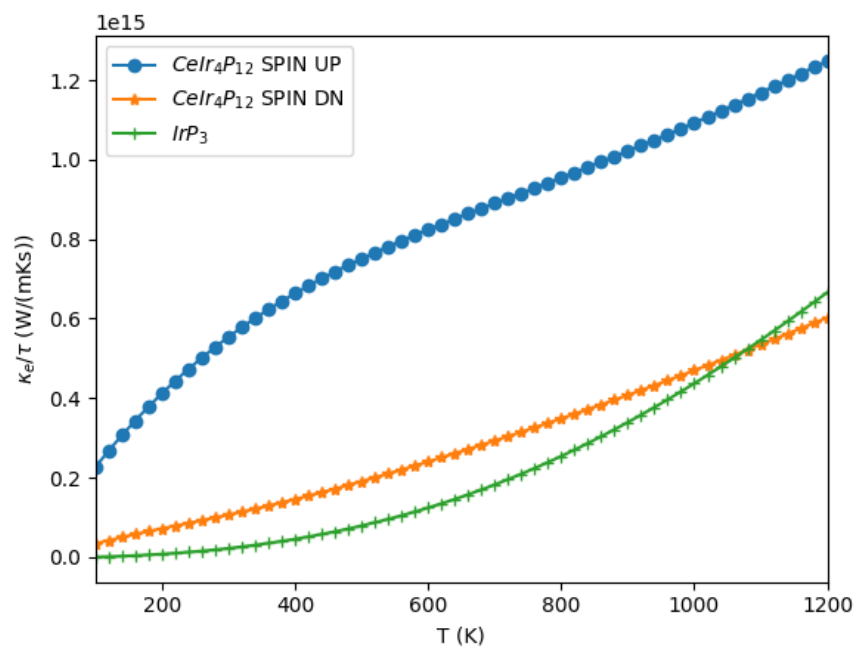}
    \caption{(Colour online) Electronic thermal conductivity ($\kappa_e/\tau$) in W(mK~s)$^{-1}$  as a function of temperature for $\textrm{Ir}\textrm{P}_\textrm{3}$ and $\textrm{CeIr}_\textrm{4}\textrm{P}_\textrm{12}$ spin up and spin down.}
    \label{figure11}
\end{figure}

In any material the conduction is caused by lattice vibrations and electrons. The equation $\kappa=\kappa_e+\kappa_l$ indicates that lattice vibrations are responsible for heat reciprocation while free electrons in metals cause thermal conductivity. Low thermal conductivity per relaxation time ($\kappa/\tau$) leads to a better thermoelectric performance of a compound. The variation of electronic thermal conductivity $\kappa_e/\tau$ for $\textrm{Ir}\textrm{P}_\textrm{3}$ and $\textrm{CeIr}_\textrm{4}\textrm{P}_\textrm{12}$ compounds is shown in figure~\ref{figure11}. It is clear from the plot that the thermal conductivity increases with an increase in
temperature.

\begin{figure}[ht]
    \centering
    \includegraphics[width=0.55\linewidth]{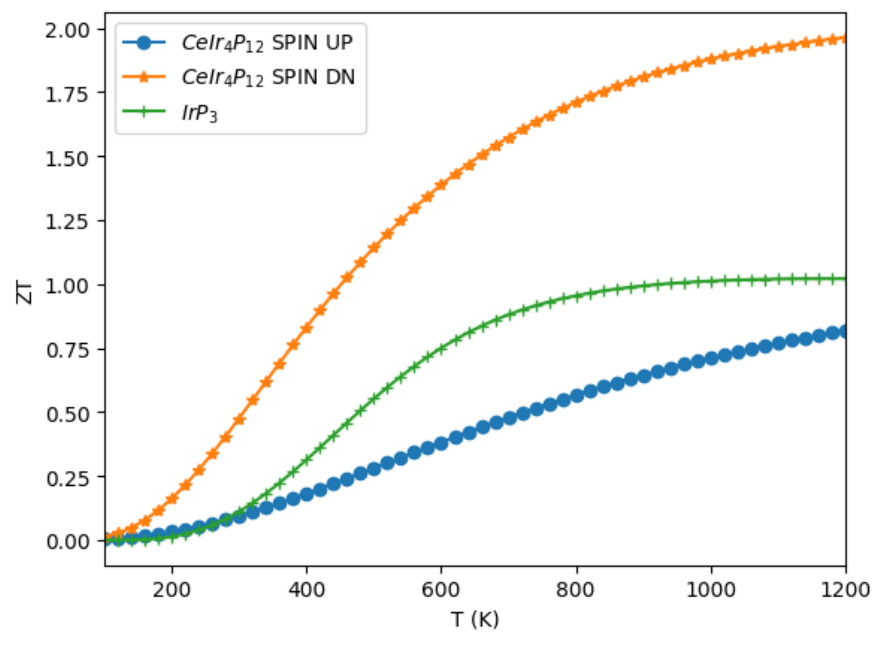}
    \caption{(Colour online) Figure of merit ($ZT$) as a function of temperature for $\textrm{Ir}\textrm{P}_\textrm{3}$ and $\textrm{CeIr}_\textrm{4}\textrm{P}_\textrm{12}$ spin up and spin down.}
    \label{figure12}
\end{figure}

The thermoelectric performance is characterized by dimensionless figure of merit $ZT$ defined as:
\begin{equation}
   ZT=\frac{S^2\sigma}{\kappa_e+\kappa_l}T .
\end{equation}
After calculating all the transport parameters, Seebeck coefficient ($S$), electrical conductivity ($\sigma$) and thermal conductivity ($\kappa_e+\kappa_l$), one can anticipate the thermoelectric efficiency ($ZT$) of the compound.
The value of the figure of merit ($ZT$) for binary skutterudite increases with an increase in temperature. At 1200 K, the obtained value of the figure of merit is 1.02. For the $\textrm{CeIr}_\textrm{4}\textrm{P}_\textrm{12}$,  $ZT$ increases with the temperature to reach its maximum at 1200 K. In spin up channel at 1200 K, the value of $ZT$ is 0.82. At the same temperature in spin down channel, the value of $ZT$ is 1.96 (see figure \ref{figure12}). A comparison of the figure of merit of the Ce-filled ternary skutterudite shows that its figure of merit increases in the spin down channel but decreases in the spin up channel.

\section{Conclusion}
\label{section4}
In summary, we have studied the structural, electronic, elastic, and transport properties of $\textrm{Ir}\textrm{P}_\textrm{3}$ and predicted a rare earth filled skutterudite $\textrm{CeIr}_\textrm{4}\textrm{P}_\textrm{12}$ by using the full-potential linearized augmented plane wave (FP-LAPW) method as employed in WIEN2k code. The exchange-correlation energy was obtained using generalized gradient approximation (GGA) by Wu and Cohen (GGA-WC) and local density approximation (LDA). In case of binary compound, the lattice constant is in good agreement with the available theoretical as well as experimental results. The predictive range of experimental lattice parameter of $\textrm{CeIr}_\textrm{4}\textrm{P}_\textrm{12}$ is [8.197, 8.238]. We have found that this compound is thermally stable from point of the energy of formation. This predicted skutterudite compound ($\textrm{CeIr}_\textrm{4}\textrm{P}_\textrm{12}$) is also elastically stable. The compound is found to be ductile in nature and anisotropic. Moreover, the compound verified the dynamic stability criterion where we found that all the frequencies of the acoustic and optic phonon branches were positive. The obtained electronic results show that $\textrm{Ir}\textrm{P}_\textrm{3}$ is semiconductor with indirect narrow band gap. The spin-polarized band structure and density of states of $\textrm{CeIr}_\textrm{4}\textrm{P}_\textrm{12}$ show its behavior as metallic. The Hubbard parameter ($U$) has been used with GGA-WC functional to see the effect on band gap between valence band and the conduction band of ternary skutterudite. The behavior of macroscopic thermodynamic parameters with temperature was predicted using the quasi-harmonic Debye-Slater model. Thermoelectric investigations indicate high $ZT$ values (1.02, 0.82 and 1.96 for $\textrm{Ir}\textrm{P}_\textrm{3}$, $\textrm{CeIr}_\textrm{4}\textrm{P}_\textrm{12}$ spin up channel and $\textrm{CeIr}_\textrm{4}\textrm{P}_\textrm{12}$ spin down channel, respectively) due to the high Seebeck coefficient and low total thermal conductivity at 1200 K. The findings of the present manuscript open up the possibility of exploring rare earth-filled skutterudites materials in order  to fabricate high-performance
thermoelectric generators. This Ce-filled ternary compound was not studied earlier; therefore, a comparison of the results is not given.

\section*{Acknowledgements}
Thanks the financial support obtained from University of Tlemcen through project, PRFU B00L02UN130120230001, Professor Tarik Ouahrani from ESSAT Tlemcen, Algeria and Professor Smaine Bekhechi form university of Tlemcen, Algeria.

\bibliographystyle{cmpj}
\bibliography{cmpj_bib}

%
%

\ukrainianpart

\title{Теоретичне дослідження термоелектричних властивостей $\textrm{CeIr}_\textrm{4} \textrm{P}_\textrm{12}$ наповненого скуттерудиту для перетворення енергії}
\author{М. Бученакі, Л. І. Карузане, Б. Н. Брамі, М. Каід Слімане}

\address{Лабораторія теоретичної фізики, Природничий факультет університету Абу Бекр Белкаіда, Тлемсен, Алжир}

\makeukrtitle

\begin{abstract}
	\tolerance=3000%
	Структурні, пружні, термодинамічні та термоелектричні характеристики $\textrm{CeIr}_\textrm{4} \textrm{P}_\textrm{12}$ скуттерудиту розраховані на основі теорії функціоналу густини та методу напівкласичного моделювання Больцмана.  Структурно-магнітна стійкість системи була перевірена за допомогою розрахунку енергії основного стану, отриманої в результаті структурної оптимізації. Розраховані константи пружності монокристалів (${C}_{ij}$) підтверджують, що згадана сполука є механічно стійкою. Крім того, вона є динамічно стійкою, коли всі фононні частоти є додатними. Розраховано енергію когезії для перевірки енергетичної стійкості матеріалу. Також досліджено температурні залежності деяких макроскопічних фізичних характеристик, а саме: коефіцієнта теплового розширення і теплопровідності ґратки. Крім того, проведено дослідження температурних залежностей деяких термоелектричних коефіцієнтів, таких як електронна теплопровідність та добротність. Отримані результати вказують на те, що подібні сполуки можна використовувати в термоелектричних пристроях.
	\keywords метод функціоналу густини, фізичні властивості, скуттерудити, властивості переносу, числові розрахунки, термоелектричні властивості
\end{abstract}
\lastpage
\end{document}